\documentclass[figures,cite]{epl}
\usepackage{epsfig}

\title{Disorder-induced melting of the charge order in
thin films of Pr$_{0.5}$Ca$_{0.5}$MnO$_3$}
 \shorttitle{Strain relaxation induced melting }
\author{Z. Q. Yang\inst{1} \and R. W. A.  Hendrikx\inst{1}
\and P. J. M. v. Bentum\inst{2} \and J. Aarts\inst{1}}
 \shortauthor{Z. Q. Yang \etal}
 \institute{
 \inst{1} Kamerlingh Onnes Laboratory, Leiden University, P.O. Box 9504, Leiden,
 the Netherlands\\
  \inst{2} Nijmegen High Field Magnet Laboratory, Toernooiveld 1,
 6525 ED Nijmegen, the Netherlands }


\begin{document}

\maketitle \today \\

\begin{abstract}
We have studied the magnetic-field-induced melting of the charge
order in thin films of Pr$_{0.5}$Ca$_{0.5}$MnO$_3$ (PCMO) films on
SrTiO$_3$ (STO) by X-ray diffraction, magnetization and transport
measurement. At small thickness (25~nm) the films are under
tensile strain and the low-temperature melting fields are of the
order of 20 T or more, comparable to the bulk value. With
increasing film thickness the strain relaxes, which leads to a
strong decrease of the melting fields. For a film of 150~nm, with
in-plane and out-of-plane lattice parameters closer to the bulk
value, the melting field has reduced to 4 T at 50 K, with a strong
increase in the hysteretic behavior and also an increasing
fraction of ferromagnetic material. Strain relaxation by growth on
a template of YBa$_2$Cu$_3$O$_{7-\delta}$ or by post-annealing
yields similar results with an even stronger reduction of the
melting field. Apparently, strained films behave bulk-like.
Relaxation leads to increasing suppression of the CO state,
presumably due to atomic scale disorder produced by the relaxation
process.
\end{abstract}

\section{Introduction}
The occurrence of Charge Order (CO) in doped perovskite manganites
of type {\it RE}$_{1-x}${\it A}$_x$MnO$_3$ ({\it RE} = trivalent
rare earth ,{\it A'}= divalent alkaline earth) is currently a much
studied phenomenon. The CO state, a long range ordering of the
Mn$^{3+}$ and Mn$^{4+}$ ions, is the result of a complicated
competition between Coulomb interactions (between the charges),
exchange interactions (between the Mn moments), and the
electron-lattice coupling. It is therefore sensitive to the amount
of doping and to the details of the structure, but also to
magnetic fields : the insulating CO state can 'melt' into a
metallic state by polarizing the Mn moments and promoting the
mobility of the $e_g$ electron on the Mn$^{3+}$-sites. This
magnetic-field-driven insulator-metal transition leads to
'Colossal' magnetoresistance effects \cite{tokura96}. A much
studied and quite robust CO-system is Pr$_{0.5}$Ca$_{0.5}$MnO$_3$.
In the bulk, charge order sets in at 240~K, accompanied by orbital
ordering of the $e_g$-orbitals \cite{kaji01} and an increased
distortion of the orthorhombic unit cell \cite{damay98}. The
melting is hysteretic, with field values at low temperatures of
about 27~T (increasing field)and 20~T (decreasing field)
\cite{tokunaga98}. In thin film form, the development and
stability of the CO state has been much less studied. A special
issue concerns the effects of strain. Given the strong
electron-lattice coupling, it can be expected that strained films
show properties different from the bulk materials. This is the
case, for instance, in tensile strained films of
La$_{0.73}$Ca$_{0.27}$MnO$_3$ on SrTiO$_3$ (STO), where very thin
($\approx 5~nm$) films show a Jahn-Teller-like deformed structure,
and are insulating rather than metallic \cite{zand99}. Strain
release in thicker films then brings back the bulk properties.
Strain should also be present in Pr$_{0.5}$Ca$_{0.5}$MnO$_3$
(pseudocubic lattice parameter {\it a}~= 0.381~nm) grown on STO
({\it a}~= 0.391~nm). Recently reported results on this
combination demonstrated strongly reduced melting fields
\cite{prellier00a,prellier00b} for films in a thickness range
75~nm - 100~nm, which was ascribed to the fact that the
distortions normally induced by the CO state cannot fully develop
due to the strain imposed by the substrate. \\
 In the present work, we report on a similar study on
Pr$_{0.5}$Ca$_{0.5}$MnO$_3$ (PCMO) thin films of varying
thickness, deposited on STO-[100] by dc magnetron sputtering, but
we come to a different conclusion. At small thickness (25~nm) the
strained films still require high CO melting fields $H_m$ of the
order of 20 T, quite close to the value of bulk single crystals
\cite{tokunaga98}. With increasing film thickness, the strain
relaxes but the bulk-like behavior is increasingly lost; still, in
the thickness range around 80~nm, $H_m$ is significantly higher
than found in refs~\cite{prellier00a,prellier00b}. At thicknesses
around 150~nm the films are almost free of strain and $H_m$ at
50~K has reduced to 4~T, with a strong increase in the hysteretic
behavior and the appearance of a ferromagnetic signal. The data
suggest that the strain itself does not impede formation of the CO
state, but that the relaxation leads to the observed reduction of
$H_m$, presumably due to the generation of lattice defects. This
conclusion is supported by the behavior of films which are
post-annealed or grown on YBa$_2$Cu$_3$O$_{7-\delta}$ (YBCO) as
template layer : such films are more relaxed than when grown
directly on STO and shows correspondingly smaller values for
$H_m$.

\section{Experimental}

All films studied were sputter deposited from ceramic targets of
nominally Pr$_{0.5}$Ca$_{0.5}$MnO$_3$ and YBa$_2$Cu$_3$O$_7$ on
STO substrates, in a pure oxygen atmosphere of 300 Pa with a
substrate-source on-axis geometry. The high pressure leads to a
very low growth rate of 0.4 nm/min and 2.5 nm/min for PCMO and
YBCO respectively. Bilayers were grown by rotating the sample from
one target position to the other. The growth temperature was
chosen at 840$^{\circ}$C, in order to be able to grow high-quality
films of both materials at identical condition. The samples were
cooled to room temperature after deposition without
post-annealing, which leads to non-superconducting
YBCO$_{7-\delta}$ with $\delta$ =~0.53 (as determined from the
lattice parameter). Magnetotransport measurements up to 9~T were
performed with an automated measurement platform (called PPMS);
magnetization up to 5~T was measured with a SQUID-based
magnetometer (both from Quantum Design). Measurements in fields
above 9~T were performed in a Bitter magnet at the High Field
Magnet Laboratory (Nijmegen). The crystal structure and lattice
parameters were characterized by X-ray diffraction, with the
lattice parameters out-of-plane determined from the (010)$_c$,
(020)$_c$ and (030)$_c$ reflections ($c$ refers to the pseudocubic
cell, with the $b$-axis taken perpendicular to the substrate), and
in-plane from the (013)$_c$ and (023)$_c$ reflections.

\section{Results and discussion}
The structure of bulk PCMO is orthorhombic ({\it Pnma}) with {\it
a}~= 0.5395~nm, {\it b}~= 0.7612~nm and {\it c} = 0.5403~nm
\cite{Jirak85}. In terms of a pseudocubic lattice parameter $a_c$,
this means a slight difference between the a-c plane ($a_c$ =
0.3818~nm) and the b-axis ($a_c$ = 0.3806~nm). Electron
diffraction to determine the film orientation showed that for thin
films (below roughly 80~nm) the [010] axis of the film is
perpendicular to the substrate, in accordance with the findings of
ref.~\cite{prellier00a}. For thick films ($\approx 150$~nm) the
preferential orientation is the same, but domains with the
[010]-axis in the substrate plane are also found.
\begin{figure}
\onefigure{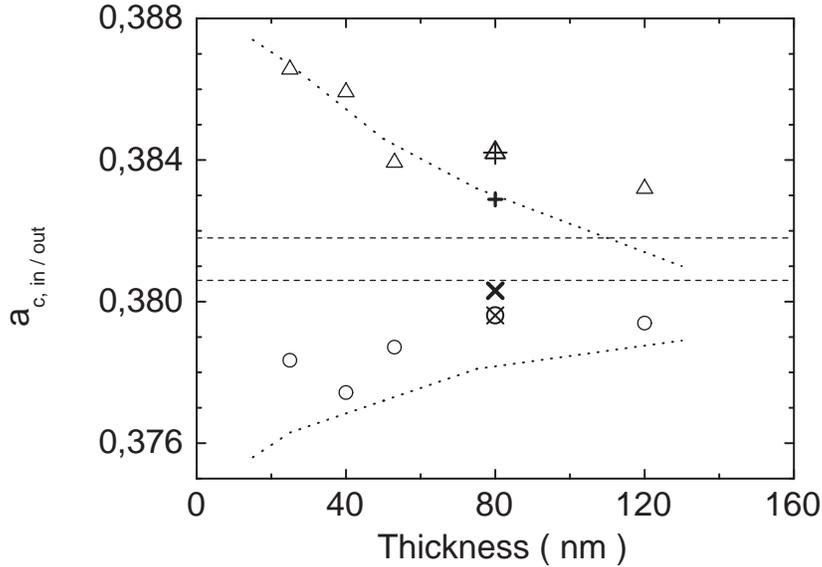}
 \caption{Lattice parameters
($\circ$ : out-of-plane $a_{c,out}$; $\triangle$ : in-plane
$a_{c,in}$) for films of Pr$_{0.5}$Ca$_{0.5}$MnO$_3$ with
different thickness. The dotted lines show the behavior for
$a_{c,in, out}$ as found in ref.~\cite{prellier00a}. The
horizontal dashed lines indicate the bulk values.The symbols
circle/plus and triangle/cross denote a 1-hour post-annealed film
of 80~nm; (+,x) denote the same film after a 5-h post-anneal.}
 \label{fig:f1-latpar}
\end{figure}
The thickness dependence of in-plane and out-of-plane lattice
parameters $a_{c,in,out}$ is plotted in Fig.~\ref{fig:f1-latpar}.
At low thickness $a_{c,in}$ is closer to the (larger) substrate
value than to the bulk value, while $a_{c,out}$ is smaller than
the bulk value, indicating that the films grow epitaxially and
strained. With increasing thickness both lattice parameters tend
towards the bulk values. The behavior is quite similar to that
reported in ref.~\cite{prellier00a} as indicated by the dotted
lines in Fig.~\ref{fig:f1-latpar}. The full-width-at-half-maximum
of the rocking curve of the (020) peak for all films is smaller
than 0.5$^{\circ}$, indicating good crystallinity. \\

\begin{figure}
\onefigure{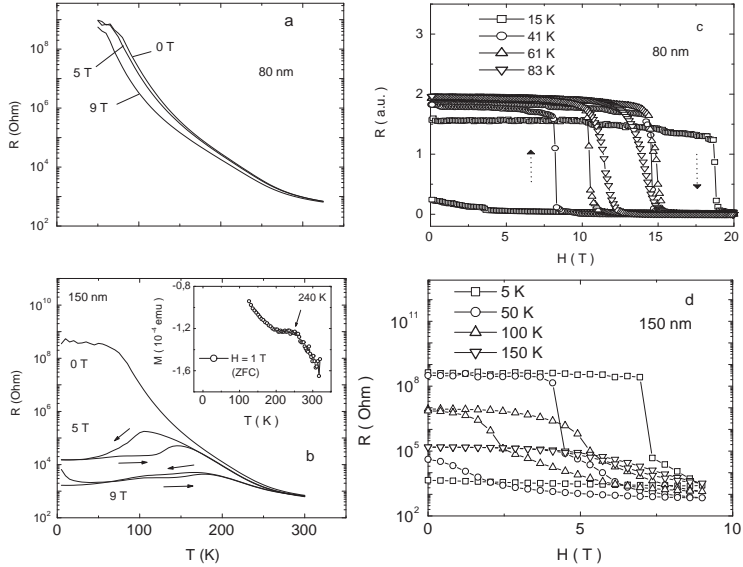} \caption{Resistance $R$ versus
temperature $T$ at magnetic fields $H$ = 0, 5, 9~T for films of
Pr$_{0.5}$Ca$_{0.5}$MnO$_3$ with thickness (a) 80 nm, (b) 150 nm;
for the same films $R$ versus $H$ at different $T$ as indicated,
(c) 80~nm, (d) 150~nm } \label{fig:f2-RTH}
\end{figure}

All films showed semiconductor-like insulating behavior in zero
applied magnetic field, as illustrated in Fig.~\ref{fig:f2-RTH}a,b
for films of 80~nm and 150~nm. An anomaly is present in the
logarithmic derivative $dR/d(1/T)$ around a value expected for the
CO transition temperature $T_{co}$, but without the jump which is
prominently observed in bulk material at 240~K. The absence of
this jump is probably due to the fact that the increase of the
in-plane lattice parameter which accompanies the charge ordering
is already accomodated by the substrate strain \cite{ogim01}. The
CO transition is visible in the magnetization $M$, especially for
the thicker films. As shown in the inset of
Fig.~\ref{fig:f2-RTH}b, $M(T)$ for the 150~nm film in a field of
1~T shows a clear shoulder around 240~K, reminiscent of the peak
in the susceptibility found in the bulk material at $T_{co}$ (and
above the magnetic transition) \cite{Jirak00}.

\begin{figure}[t]
\onefigure{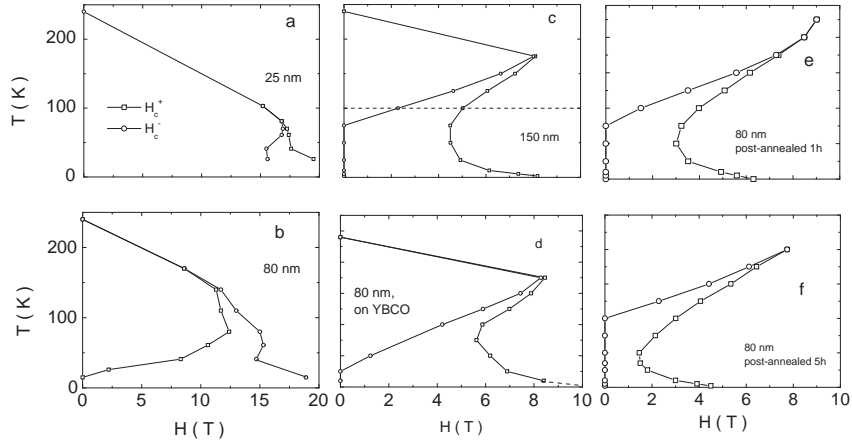}
 \caption{Charge order melting field phase diagrams as determined from
the magnetoresistance for films of different thickness. The point
at zero field is the bulk value for $T_{co}$. (a) 25~nm; (b)
80~nm; (c) 150~nm; (d) 80~nm, grown on YBCO template; (e) 80~nm
post-annealed 1h; (f) 80~nm post-annealed 5h. The dashed line in
(c) denotes the temperature of the magnetization measurements
given in Fig.~\ref{fig:f4-mhloop} }
 \label{fig:f3-phasedia}
 \end{figure}

For films of 25~nm, the resistance drop in a magnetic field, which
signifies the CO melting, was found at temperatures below 100~K
near the maximum available field of 20~T for one sample, while a
second one did not show a change in $R$ up to 20~T. Also, $R(H)$
is hysteretic~: upon decreasing the field the resistance jumps
back up at a lower field, as expected since the melting transition
is first order. We denote the upper and lower critical fields as
$H_c^+$ and $H_c^-$ respectively. With increasing thickness, both
branches shift to lower fields. Examples of $R(H)$ for the films
of 80~nm and 150~nm are given in Fig.~\ref{fig:f2-RTH}c,d. For the
film of 150~nm the melting field has dropped to only 5~T around
50~K.
The resistance changes are sharp, making $H_c$ well-defined, and
the curves can be used to construct the temperature-field phase
diagrams \cite{tomioka96} shown in Fig~\ref{fig:f3-phasedia},
where at zero field the value of the bulk is used.
The shape of the phase diagrams changes significantly with
increasing thickness. For the 25~nm film, hysteresis is only
present below 70~K in the field region 16~T - 20~T. These large
values resemble the numbers found for bulk single crystals. For
the 80~nm film hysteresis starts below 130~K and the difference
between the ($+,-$)-branches increase considerably, especially at
low temperatures; the $H_c^+$ branch is still above 12~T at all
temperatures, which explains the small MR effects seen in
Fig.~\ref{fig:f2-RTH}a. Both $H_c^+$ and $H_c^-$ are considerably
larger than reported in ref~\cite{prellier00b}. In the 150~nm film
hysteresis is found below 175~K. Both branches have shifted to
lower fields~: $H_c^+$ is curved with a minimum value of 4~T
around 50~K, while $H_c^-$ now lies at zero field for temperatures
below 80~K. \\

\begin{figure}
\onefigure{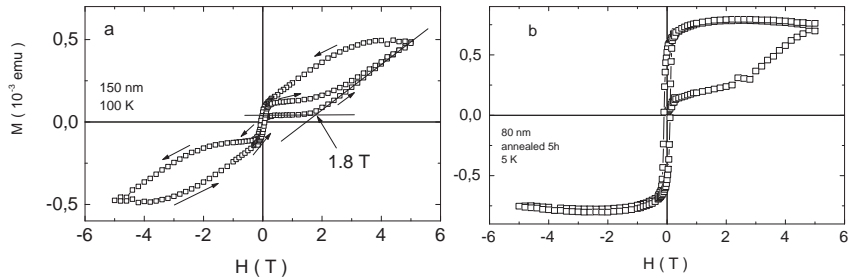}
 \caption{Magnetization $M$ versus magnetic field $H$ for films of
Pr$_{0.5}$Ca$_{0.5}$MnO$_3$. In both cases, the magnetization of
the substrate has been subtracted. (a) film of 80~nm at 100~K; (b)
film of 80~nm, post-annealed for 5h, at 5~K.}
 \label{fig:f4-mhloop}
\end{figure}

The melting transition is insulator-metal, but also
antiferromagnetic-ferromagnetic, and can therefore be seen in the
field dependence of the magnetization. Fig.~\ref{fig:f4-mhloop}a
shows $M(H)$ of the 150~nm film at 100~K, for the field sequence
0~T $\rightarrow$ +5~T $\rightarrow$ -5~T $\rightarrow$ +5~T, with
the diamagnetic substrate signal subtracted. A small ferromagnetic
component is already present in the virgin state; with increasing
field $M(H)$ is constant until 1.8~T and then rises significantly
when the field is increased to 5~T. Upon decreasing the field $M$
now remains constant because the sample is in the FM state, but
starts to drop around 3~T when $H_c^-$ is crossed as can be seen
in Fig~\ref{fig:f3-phasedia}c (dotted line). At zero field, the
ferromagnetic component has grown by more than a factor 2. The
same behavior is found when continuing the loop to -5~T; when
going back up to +5~T, $M$ merges with the virgin curve above 4~T.
\\

The first conclusion we draw is that the CO state in the strained
material is hardly less stable (if at all) than in the bulk. This
is different from the one reached in
Refs.~\cite{prellier00a,prellier00b}, but it is in good agreement
with the data reported on Cr-doped films \cite{ogim01} : in that
case it was found that strain-free films very quickly developed
ferromagnetism upon Cr-doping, but that Cr-doped films under
tensile strain were still insulating, suggesting that the strain
counteracts the effects of the Cr doping and stabilizes the CO
state. The decreasing stability of the CO state with increasing
film thickness appears due to the strain relaxation rather than
the strain itself. The picture arising then is that defects
(disorder) induced by the growth and the relaxation destabilize
CO, but that the strain itself has no destabilizing effect or even
the opposite, which is quite reasonable in view of the fact that
the necessary lattice distortion is already accommodated (also
suggested in ref.~\cite{ogim01}). \\

In order to highlight the effects of strain relaxation we
performed two more experiments. One 80~nm film was annealed in the
growth chamber for one hour at 950 $^{\circ}$C in 1~mbar O$_2$
(the sputtering pressure) and slowly cooled; after measuring it
was annealed for an additional 5 hours in flowing oxygen at 900
$^{\circ}$C. Another 80~nm film was grown on a 10~nm YBCO template
layer (called PY) since in previous work \cite{yang01} we found
that YBCO is an effective strain relaxor for
La$_{0.67}$Ca$_{0.33}$MnO$_3$. Both methods effectively relax the
strain in PCMO as well. Lattice parameter values ($a_{c,out}$,
$a_{c,in}$) are (0.384~nm, 0.380~nm) for the 1-hour post-annealed
sample, (0.383~nm, 0.380~nm) for the 5-hour post-annealed sample ,
and (0.385~nm, 0.380~nm) for PY, showing that all have undergone
relaxion, especially in the out-of-plane axis. The CO-melting
phase diagrams for these samples again show a strong decrease of
the melting fields (see Fig.~\ref{fig:f3-phasedia}d-f), with the
5-hour post-annealed sample reaching the lowest value yet observed
in this system (1.5~ T at about 50~K). The field dependence of the
magnetization e.g. at 5~K (Fig.~\ref{fig:f4-mhloop}b) accordingly
shows an increase of $M$ around 3~T (due to the bending back of
the $H_c^+$-branch), but no decrease of $M$ from 5~T downward
until the ferromagnetic hysteresis regime is entered, since
$H_c^-$ now lies at 0~T. \\

Since relaxation by post-annealing must be accompanied by inducing
defects in the film, the observations reinforce the notion that
defects are responsible for the change in melting behavior. In
this respect it is important to note that the development of the
phase diagrams presented in Fig.~\ref{fig:f3-phasedia} closely
resembles the changes found in the bulk material when going from
$x$~= 0.5 (small hysteretic regime at a large field) to $x$~=0.3
(curved upper branch at relatively low fields and lower branch
going to zero) \cite{tokura96} ; especially the similarity between
the behavior of the 5h~post-annealed film and the $x$~=0.3 bulk
material is striking, with both showing a minimum $H_c^+$ field of
about 2~T around 30~- 40~K, and the $H_c^-$-branch at zero field.
Still, the physics behind this may not be quite the same. In the
bulk case, the change of doping induces discommensurations and a
canted antiferromagnetic (c-a-f) state. In the films the amount of
carriers is not changed; rather it is the local structure which
can vary, which would influence the local Jahn-Teller distortions
and the concommittant orbital order. This would in turn promote
ferromagnetic interactions, possibly leading to ferromagnetic
clusters in a phase-separation-like scenario very similar to the
disorder-driven phase separation observed in films of
La$_{0.67}$Ca$_{0.33}$MnO$_3$ \cite{biswas00}. We observe that the
structure relaxation is accompanied by an increasing amount of
ferromagnetic component in the magnetization, which could be
either due to the c-a-f state or to ferromagnetic clusters. The
answer to this question may come from electron microscopy studies,
which are now in progress. Finally, we note that disorder as a
major source for reduced melting fields can also explain the
difference between our results and those of
ref.~\cite{prellier00a,prellier00b} as caused by the different
morphology of the sputtered versus the laser-ablated films. This
suggests that the CO state is more sensitive to disorder than
might be assumed in view of the high melting fields, and that if
CO films are to be grown, avoiding disorder is the major source of
concern.
\\
\\
In summary, we have shown that the melting fields $H_m$ for the
insulating CO state in Pr$_{0.5}$Ca$_{0.5}$MnO$_3$ films under
tensile strain are around 20~T or even above, rather close to the
values found for the bulk material. Strain relaxation strongly
reduces $H_m$. With increasing film thickness, the lattice
parameters of the film demonstrate relaxation, while $H_m$
decreases down to 4 T at 50~K for a 150-nm film. Upon strain
relaxation by post-annealing this value becomes even smaller. We
suggest this is due to induced defects, which destabilize the
antiferromagnetic state and possibly even promote the formation of
ferromagnetic clusters.

\acknowledgments
 This work is part of the research program of the
'Stichting voor Fundamenteel Onderzoek der Materie (FOM)', which
is financially supported by NWO.

\end{document}